# Combining distributive ethics and causal Inference to make trade-offs between austerity and population health

Adel Daoud[1, 2, 3*], Anders Herlitz[4, 5], and S V Subramanian[1,6]

8 August, 2020

1. Center for Population and Development Studies, Harvard T.H. Chan School of Public Health, Harvard University, Boston MA, USA.
2. Department of Sociology and Work Science, University of Gothenburg, Sweden.
3 The Division of Data Science and Artificial Intelligence, the Department of Computer Science .
4. Department of Global Health and Population, Harvard T.H. Chan School of Public Health, Harvard University, USA.
5. Institute for Futures Studies, Stockholm, Sweden.
6. Department of Social and Behavioral Sciences, Harvard T.H. Chan School of Public Health, Harvard University, USA

* Corresponding author: adaoud@hsph.harvard.edu

# Contents

**Abstract**: The International Monetary Fund (IMF) provides financial assistance to its member-countries in economic turmoil, but requires at the same time that these countries reform their public policies. In several contexts, these reforms are at odds with population health. While researchers have empirically analyzed the consequences of these reforms on health, no analysis exist on identifying fair tradeoffs between consequences on population health and economic outcomes. Our article analyzes and identifies the principles governing these tradeoffs. First, this article reviews existing policy-evaluation studies, which show, on balance, that IMF policies frequently cause adverse effects on children's health and material standards in the pursuit of macroeconmic improvement. Second, this article discusses four theories in distributive ethics (maximization, egalitarianianism, prioritarianiasm, and sufficientarianism) to identify which is the most compatible with IMF's core mission of improved macroeconomics (Articles of Agreement) while at the same time balancing consequences on health. Using a distributive-ethics analyses of IMF polices, we argue that sufficientarianism is the most compatible theory. Third, this article offer a qualitative rearticulation of the IMF's Articles of Agreement, and formalize sufficientarian principles in the language of causal inference. We also offer a framework on how to empirically measure, from observational data, the extent that IMF policies trade off fairly between population health and economic outcomes. We conclude with policy recommendations and suggestions for future research.

# Introduction

The International Monetary Fund (IMF) has at its mission to monitor global financial systems and provide macroeconomic support to its 189 member-countries. It conditions this support on its own policies, the harshest being structural adjustments and austerity. Given the often invasive nature of IMF policies, subsequent consequences on population health programs have spurred considerable debates (Babb 2005; McKee et al. 2012; Stuckler and Basu 2013; Summers and Pritchett 1993). Researchers have conducted empirical evaluations of IMF policies to identify which affect population health (Clements, Gupta, and Nozaki 2013; Daoud et al. 2017; Daoud and Reinsberg 2018; Dreher 2009; Kentikelenis, Stubbs, and King 2016; Shandra, Shandra, and London 2012; Stubbs, Kentikelenis, and King 2016; Vreeland 2007).

However, these empirical studies lack a key component found in distributive ethic studies; that is, transparency in what is considered fair and desirable outcomes. Ethics-based studies use normative arguments that focus on *ought* issues—they question and evaluate desirable outcomes—while empirical studies favor explanatory or descriptive arguments that focus on *is* issues—they evaluate causes and effects. Leaving out ought issues, empirical studies of IMF policies do not say much about how the IMF should trade off between population-health outcomes and macroeconomic outcomes.

A distributive-ethics analysis of the Articles of Agreement that govern the IMF's mission entails identifying which principles would reflect a fair tradeoff between health and economic outcomes (Daniels 1985). This trade-off pertains to the distribution of outcomes in mainly two dimensions, or outcomes: *wellness of macroeconomics* and *wellness of individuals*. The former refers to all outcomes that are considered beneficial for a population's economy in aggregate (e.g., higher economic growth, more trade, stable currency); the latter refers to all outcomes that are considered beneficial for a person as an atomic unit (e.g., better physical and mental health, less stress, access to sufficient material assets). While these two outcomes often reinforce each other, on several occasions—such as in the context of structural adjustments or austerity—they can be at odds with each other (McKee et al. 2012; Stuckler and Basu 2013). For example, when the IMF balances a country's budget, its austerity policies tend to stipulate less spending on health and social systems.

Our article (a) reviews the findings of empirical studies focusing on what (causal) direction IMF policies push the wellness of individuals, focusing on children's health as a case study; (b) evaluates which of four theories in distributive ethics is appropriate for governing IMF Articles of Agreement while minimizing potential harm to individuals' outcomes; and (c) formalizes this distributive-ethics in causal-inference language. The four theories are maximization, egalitarianianism, prioritarianiasm, and sufficientarianism (section 3 defines these theories).

We present our arguments in four steps. Section I discusses the IMF's Articles of Agreement and how they determine the IMF's focus on the wellness of macroeconomics. Section II reviews the findings of applied policy evaluation studies, which show that in many cases IMF policies cause adverse effects on children's health and material standards. We offer a summary that supplies an empirical foundation for the real relevance of conducting a distributive-ethics analysis of IMF policies. Section III evaluates four theories in distributive ethics and identifies which is the most compatible with balancing the IMF's core mission on macroeconomics with individual well-being. Having made our argument in favor of sufficientarianism, Section IV proposes a qualitative rearticulation of the IMF's Articles of Agreement and a formalization of these thoughts. We conclude this article with suggestions for future research.

# Section I: Evaluating implications for health in IMF Articles of Agreement

The IMF operates within the framework established by its 189 member-countries. The parameters of this framework are defined by the IMF's Articles of Agreement, which were adopted at the United Nations Monetary and Financial Conference, in Bretton Woods, New Hampshire, on July 22, 1944. These articles have been amended seven times, the last amendment being adopted on December 15, 2010 (IMF 2011). A critical part of this agreement is Article V, section 3(a), because it shapes IMF's focus on the wellness of macroeconomics (Vreeland 2007):

> The Fund shall adopt policies on the use of its general resources … that will assist members to solve their balance of payments problems in a manner consistent with the provisions of this Agreement and that will establish adequate safeguards for the temporary use of the general resources of the Fund. (IMF 2011)

This article binds the IMF to assist its member-countries, but only in solving the balance of payment problems. Accordingly, this and reinforcing articles bind the IMF to push for public policies in borrowing countries that improve the wellness of macroeconomics (Clements et al. 2013; IMF 2016) but that simultaneously de-prioritize other issues such as population health and poverty.

While Article V focuses the IMF on correcting macroeconomic fundamentals, Article I includes that these corrections should not damage national prosperity. The IMF is allowed to supply resources "without resorting to measures destructive of national or international prosperity" (IMF 2011). Nonetheless, as evident by both policy and scientific debates, IMF's measures have not been adequate (Babb 2005; Bohoslavsky 2016; Stiglitz 2003; Summers and Pritchett 1993). In response to this critique, the IMF has sought to incorporate several social policies into its programs (IMF 2014). By the mid-1990s, the IMF introduced policies designed to protect poor populations via so-called Poverty Reduction Strategies (Gupta et al. 2000). These strategies ranged from setting a minimum amount that a country should spend on health, known as priority expenditure floors (de Rato 2006), to implementing targeted social safety nets (Clegg 2014). Low-income countries, which are the most common borrowers, are given access to concessionary funds (Barro and Lee 2005).

Despite IMF's effort to reduce poverty, several empirical studies signal that these efforts might still be insufficient to counter the adverse effects on individual well-being via other causal channels—such as from other economic policies bundled into the same program. Between 1985 and 2014, the IMF operated 1,550 programs in over 130 countries (Kentikelenis et al. 2016). In these different programs, the IMF launched a total of 55,465 policies, but only 822 (1.5 percent) of these were explicitly aimed to reduce poverty. The vast majority (87 percent) were designed to address macroeconomics imbalances (Kentikelenis et al. 2016). An IMF program has, on average, 36 policies. Assuming that social policies are uniformly distributed, each program—on average—contained only 0.36 (that is, less than one) social policy.

The IMF's Articles of Agreement match well with what the organization does empirically: it prioritizes macroeconomic issues over individual well-being (Copelovitch 2010; Dreher 2006). This prioritization implies that even if IMF officials have an interest in improving the wellness of individuals, the Articles of Agreement tie them to prioritize macroeconomic over individual well-being. In the next section, we evaluate the empirical evidence of how this prioritization has affected children.

## Section II: Reviewing the empirical evidence on the effect of the International Monetary Fund's policies on children's well-being

Now that we reviewed how the Articles of Agreement regulate IMF policies, we proceed with evaluating the findings of empirical studies. Our question is: In what directions do IMF policies push child health and poverty? Thompson et al. (2017) recently reviewed studies on the effect of international financial organizations on maternal and child health. These organizations include the IMF, the World Bank, and the African Development Bank. Their search identified six studies of interest, focusing on the effect of IMF policies on children. We updated this search, focusing only on IMF programs, to cover the period 2017 to date. We found three additional studies. Table 1 outlines these nine studies.

[Table 1 about here]

Seven of the nine articles identify adverse effects. While four studies find adverse direct effects on different child outcomes (i.e., mortalitly, poverty, vaccination, undernutrition) (Daoud and Johansson 2019; Oliver 2006; Pongou, Salomon, and Ezzati 2006; Shandra et al. 2012), three studies find adverse indirect effects (interactions) through a third variables (i.e., democracy, education, public sector policies) on children (Daoud et al. 2017; Daoud and Reinsberg 2018; Shandra et al. 2004). Two studies identify a beneficial direct effect: Bird et al. (2020) and Hajro and Joyce (2009) each find that IMF programs reduce infant mortality.We discuss the findings of the nine studies by identifying the direct and indirect pathways in which IMF programs are likely to affect child poverty.[1] These pathways provide the foundation for our substantive argument and formalization of our ethical approach in Section IV.

---

[1] We include additional references above these nine studies where helpful.

**Direct pathways**. IMF programs affect government spending (Bird et al. 2020), as governments often subsidize the costs of immunizations, food, and health services (Daoud 2007, 2015; Halleröd et al. 2013). Given that the IMF aims to balance the borrowing-government budget, these subsidies are often the first to be dismantled (Shandra, Shandra, and London 2011). Another direct pathway through which IMF programs are likely to affect child poverty is in the restructuring of health sectors (Daoud and Reinsberg 2018; Oliver 2006; Stuckler and Basu 2013). The IMF operates under the assumption that private actors are more efficient in providing healthcare and other public services (Benson 2001). While the overall quality of health care might improve, privatization tends to favor rich households (Daoud and Johansson 2019; Ismi 2004; Pongou et al. 2006). Those that gain from privatization thus tend to be the economically privileged segments of urban populations, while those that lose tend to concentrate in remote rural areas, far removed from modern corporations' logic of private profit (Shandra et al. 2012).

IMF programs also have a catalyzing effect on foreign investments and aid (Clements et al. 2013). Recent research finds that this effect is stronger in sectors linked to the IMF's core policy areas than to its noncore areas. This could, in turn, for example, give leverage to countries to combat health deprivations (Stubbs et al. 2016:511). This means that IMF programs may reduce child poverty via the influx of such resources (Bird et al. 2020; Hajro and Joyce 2009).

**Indirect pathways.** The IMF can also affect child poverty via indirect channels, such as unintended (i.e., second-order) consequences of policy reforms. The first indirect pathway operates through state employees (Daoud and Reinsberg 2018). The IMF programs within the current framework cap public sector wages (Rickard and Caraway 2014), in a context in which a significant portion of health spending in low-income countries is allocated to the wages of doctors and other medical staff (Kentikelenis, Stubbs, and King 2015). These consequences may have deleterious impacts on the provision and quality of health services. How this may happen is explained next.

The second indirect pathway runs through international trade. The IMF currently seeks to liberalize countries' trade and investment regimes, for example by recalibrating tariffs, quotas, duties, and taxes (Dreher 2009). While these policies may eventually be beneficial for countries, they often first bring short-term lower revenues. This reduces budgets for targeted social spending, which reinforces the first direct channel outlined above (Daoud and Johansson 2019; Pongou et al. 2006). Moreover, IMF policies involve currency devaluations that foster import increases, inflate medicine and medical equipment prices, drive up the cost of clean water and sanitation equipment,[2] and reduce children's access to vaccines and other crucial health-related resources (Shandra et al. 2011).

The third indirect pathway is global politics. The IMF takes different negotiation positions based on which country it is entering an agreement with (Vreeland 2007), which in turn can affect how many resources a country gets. For example, a final agreement tends to be more favorable if the country has any substantial influence on the global economy or if it is compliant with IMF demands (Oliver 2006; Pongou et al. 2006). Emerging powers, such as Brazil and Turkey, tend to have more of a say. Sub-Saharan African countries have weaker negotiation positions—as do other low-income countries—but they tend to get different treatment because of their high poverty rates, such as favorable conditionality under IMF's Poverty Reduction Strategy or some debt forgiveness (Dreher 2009). Consequently, these contextual or geo-historical channels can moderate the strength of the relationship between an IMF program and children's well-being, independently of the factors outlined above.

---

[2] This refers to a setting in which the availability of clean and safe water is a core determinant of infant mortality (Shandra, Shandra, and London 2011).

All of these direct and indirect impacts of IMF programs on child poverty depend on household resilience (Daoud et al. 2017; Daoud and Puaca 2011; Puaca and Daoud 2011), which refers to the family members' ability to mitigate, adapt to, and recover from economic, natural, or political shocks and stresses (Conklin et al. 2018; Coutts et al. 2019; Daoud, Halleröd, and Guha-Sapir 2016; Daoud et al. 2016; Daoud and Nandy 2019; Kraamwinkel et al. 2019; Nandy, Daoud, and Gordon 2016; Ponce et al. 2017). Households with more resources tend to be more resilient than those with fewer resources. Nevertheless, household resilience itself can be affected by adjustment policies, negatively or positively. For example, changes in labor market policies (e.g., adjustments to minimum wage levels) will affect parents' income and ability to care for their children.

In sum, despite the IMF's several efforts to reduce the adverse effects of its programs on population health, the majority of the empirical research finds a net adverse effect on children. These findings, therefore, reinforce the argument that transforming the IMF's Articles of Agreement requires injecting them with elements of distributive ethics. With such an injection, the IMF has a better ethical ground to balance the wellness of macroeconomics and individuals.

## Section III: Identifying distributive ethics for amending the Articles of Agreement

We next evaluate four distributive-ethics theories: *maximization*, *egalitarianianism*, *prioritarianiasm*, and *sufficientarianism*. Foreshadowing our conclusion, we will argue that sufficientarianism is most compatible with the IMF's core mission—to reinvigorate the wellness of macroeconomics—and simultaneously eradicate harm to the wellness of individuals.

Distributive ethics is an area of ethics and political philosophy that addresses the wellness of different distributions (i.e., allocations) of goods across individuals (Anderson 1999; Crisp 2003; Tännsjö 1998; Temkin 1993). Distributive theories comprise at least two elements: the *conception* of goods and the *pattern* of distribution. Goods are often thought of in terms of material resources (Dworkin 1981; Rawls 1971), welfare (Dorsey 2012), opportunities (Roemer 1998), or capabilities (Nussbaum 2000; Sen 1992). For the purpose of our argument, all these capture aspects of the well-being of individuals and all say the same about the importance of reducing poverty and improving health as essential dimensions of wellness of individuals.

The pattern of distribution is critical for our argument. The theory of maximization (such as utilitarianism) is that whatever the relevant good, what matters is that as many goods as possible are generated (Ord 2013; Tännsjö 1998). If the relevant good is welfare, for example, then distributions ought to be evaluated in terms of how much welfare is produced. The theory of egalitarianism is that whatever the relevant good, what matters is that it is distributed equally (Cohen 1989; Temkin 1993, 2003). Again, if the relevant good is welfare, distributions ought to be evaluated in terms of how equally the welfare is distributed. The theory of prioritarianiasm is that whatever the relevant good, what matters is that the amount of priority-weighted goods be maximized, so that benefits are valued higher for beneficiaries who are worse off (Adler 2012; Fleurbaey 2015; Parfit 1997). If the relevant good is welfare, then distribution ought to be evaluated both in terms of how much welfare is produced and how the benefits are prioritized. The theory of sufficientarianism is that whatever the relevant good, what matters is that it is distributed so that beneficiaries have a sufficient amount of that good (Casal 2007; Crisp 2003; Shields 2012).

In all these theories, individuals are the main unit of analysis. These four theories often yield significantly different evaluations of alternative distributions, but common to all of them is the distribution of goods across individuals, not groups, regions, or countries. This implies that, for all these theories, the wellness of macroeconomics is relevant only if it enhances the wellness of individuals.

Focusing on distributions across individuals does not imply that the four theories are insensitive to the distribution of goods among groups of individuals. Distributive ethicists often recognize the importance of groups, but mainly as an explanatory or justifying factor for why specific patterns of distribution exist (Anderson 1999). For example, the unequal distribution of opportunity (i.e., a good) among individuals in a society such as the United States can be explained by groups (e.g., race and ethnicity). In this example, the distribution of opportunities (*explanandum*) is explained by race and ethnicity (*explanans*). All four distributive-ethical theories argue that it is misguided to evaluate only group-level distributions. Table 2 highlights this type of distribution.

[Table 2 about here]

In relation to countries, egalitarians and prioritarians favor equal distribution (Outcome 2) while maximizers favor the most goods being generated (Outcome 1), since the sum of goods in Outcome 1 is larger than in Outcome 2. Sufficientarians' opinion would depend on how sufficiency is defined. Outcome 2 seems to have an equal distribution (less variance), and the priority-weights (the added weight of benefits going to the neediest) seem to make up for the slightly fewer goods in this outcome. However, none of these theories can be used to make decisions based on outcomes without knowing how these goods are distributed within the countries, and we know nothing about that in this example.

In Table 3 we introduce information about how goods are distributed within countries. To simplify our example, Country A has three inhabitants and Country B has two inhabits, but the general point is valid regardless of actual country size. The numbers in the brackets—denoting sets—signify the amount of goods each individual possesses, and the total designates the sum of a within-country distribution.

[Table 3 about here]

In Table 3, Outcome 2 distribution appears nearly equal across the two countries, but Outcome 1 is clearly a more equitable distribution across individuals. When more is understood about Outcome 1, prioritarians, egalitarians, and maximizers would consider it a better option than Outcome 2 because it contains more goods that are more equally distributed across countries. As noted above, sufficientarians would first need to determine sufficiency before making a decision (Daoud 2007, 2010, 2011, 2017, 2018). If this sufficiency threshold is set to 30 in this example, then sufficientarians would favor Outcome 1 because more individuals would reach over the threshold in this Outcome.

This thought experiment reveals two things. The first is that country size matters—which the IMF acknowledges. The second is that a narrow focus on aggregate-units (i.e., countries) fails to account for the distribution of goods within aggregate-units—yet the IMF's focus is on aggregated economic outcomes. On the one hand, the outcome for two countries of equal size and with an equal amount of goods might still have significant inequality because of skewed distributions. On the other hand, for two countries of equal size with an unequal amount of goods—so that one is richer than the other (and there is country-level inequality)—the outcome appears relatively equal because there are no significant inequalities across the countries.

This thought experiment also reveals the risk of analyzing only country-level statistics when evaluating individual-level outcomes—a form of the ecological fallacy in which scholars use group-level information to infer conclusions on individual-level outcomes. As discussed in Section I, the IMF's focus is on the wellness of macroeconomics, which is captured in various aggregate measures (e.g., economic growth, the balance of payment). It is, of course, true that such aggregate measures often correlate with how the wellness of individuals are measured either indirectly (that is, economic growth as correlated with individual happiness) (Inglehart et al. 2008) or directly (that is, measured by the Gini coefficient or other measures of inequality) (Sen 1992). However, as we will argue in Section IV, our distributive-ethics perspective relies on causal rather than correlational thinking. From this perspective, what is relevant is what causal effects IMF policies induce on the distribution of health, and not the general association between macroeconomics performance and individuals' health. Yet, there is only one ethical theory that can balance such causal thinking without diverting the IMF from its core mission: sufficientarianism.

**Sufficientarianism**
Sufficientarianism can provide an appropriate, ethical, and minimal framework for IMF's mission that balances the wellness of individuals while keeping the IMF's focus on macroeconomics. As noted earlier, sufficientarianism is the theory that distribution is fair if all individuals in a population have a sufficient amount of goods, and distribution is unfair if one or more individuals fall short of the given threshold.

There are several reasons to favor this approach. First, regardless of which of the four distributive-ethical theories one subscribes to, there is at least one instrumental reason to also favor sufficientarian, and that is the idea of *thresholds*. From a maximizers perspective, it is expected that increasing the total amount of goods in a population ensures that each individual has at least a sufficient amount to continue contributing to the economy and community. Egalitarians who believe that more equal outcomes are desirable will also agree with thresholds because the more people reach the baseline threshold, the less inequality there will be at baseline. Prioritarians would agree with the threshold principle for both these reasons. We point out here that a minimal ethical principle for IMF's mission is the principle of thresholds (Reddy and Daoud 2020).

Second, we argue that sufficientarianism fulfills a principle of human dignity, embraced by many international organizations. Reducing inequality is important, yet this mostly follows individuals securing a minimum amount of goods to lead a decent life (Sen 1992). Having sufficient resources is a fundamental prerequisite to participate in economic, social, and political life (Anderson 1999). In improving the wellness of macroeconomics, IMF policies can be expected to increase prospects of individuals surpassing a given threshold, or at least not affect them.

Third, adopting sufficientarianism means the focus is primarily on those who fall below a given threshold; that is, the poorest groups in society. It is natural to use this theory when combating poverty and protecting the most vulnerable groups in a population. This sufficientarian focus resonates with the poverty goals that the IMF has already committed to (Clements et al. 2013; IMF 2016). Simultaneously, sufficientarianism should appeal to those who have criticized the IMF concerning the "do no harm" principle found in medical ethics (Beauchamp and Childress 2012; Feldstein and Feldstein 1998). First and foremost, doctors must avoid actions that cause suffering; by analogy, the IMF, as a doctor of macroeconomics, should thus make sure its treatments do not cause suffering for individuals.

Sufficientarians can have different goals. For example, some focus on a fixed level of welfare (Dorsey 2012); others focus on a fixed set of capabilities (Nussbaum 2000). Furthermore, there is significant disagreement among sufficientarians about how to empirically measure optimal outcomes, although there are several recent approaches in how to use global analyses. For instance, multidimensional poverty indicators might be used to more accurately track the success of an outcome (Alkire 2015; Gordon et al. 2003). While these competing sufficientarians goals provide important insights on alternative outcomes, which one may work better varies by context. Nonetheless, what matters ultimately is that as many individuals as possible reach sufficiency thresholds (Daoud 2018; Shields 2012).

## Section IV: Amending the Articles of Agreement

Now that we have identified sufficientarianism as the most suitable distributive-ethical theory for amending the IMF's Articles of Agreement, in this section we specify what this implies empirically. We formalize our approach by combining elements from causal inference (Imbens and Rubin 2015; Pearl 2009) and algorithmic fairness (Loftus et al. 2018).

Our argument is divided into two parts. The first part pertains to the *causes* (i.e., reasons) for self-selecting into IMF programs, and the second accounts for the *effects* of IMF programs on the wellness of macroeconomics and individuals. This two-part argument is depicted in our directed acyclic graph (DAG), shown in Figure 1. A DAG captures a causal system; a node indicates a causal factor; and an arrow from *A* to *B* indicates that *A* causes *B*. Mathematically, this arrow implies that conditional distribution of *B* depends on *A* in some (parametric or nonparametric) functional form, $f(\cdot)$; hence, $f(B|A)$. For simplicity, we mainly focus on probabilities denoted $P(\cdot)$; that is, $f(B|A) = P(B|A)$. The graph is acyclic, meaning that a factor cannot cause itself. For example, $A_t$ cannot affect itself at a time point $t$ but it can affect future states of itself $A_{t+1}$. In the DAGs in Figure 1, we assume no common causes (confounding). In the next subsection, we account for the case where common causes exists, and discuss how we capture causality from observational data using do-calculus.

[Figure 1 about here]

We argue that the only causal reasons the IMF accepting countries for its programs, should be based on the wellness of macroeconomics, and this largely resonates with how the IMF currently operates. The IMF's primary mission—unlike other UN organizations such as the World Bank, the World Health Organization, Food and Agricultural Organization, or UNICEF—is to oversee global financial stability, not to engage in population health or poverty reduction. While the de jure division of labor between these UN organizations exist for historical reasons (Ziring, Riggs, and Plano 2005), blurring this division of labor would inflict incompatibilities with the organizations' mission statements—but we will argue below that organizational cooperation is necessary when evaluating the causal effects of IMF programs. Formally, this macro focus is what our DAG in Figure 1 implies, and this selection in the language probability is expressed as:

$$P(IMF_t = 1|WoM_{t-1} = 0) \geq P(IMF_t = 1|WoM_{t-1} = 1)$$

Consequently, to determine whether self-selection into IMF programs is fair in a sufficientarian sense, we have to evaluate this probabilistic inequality. The probability of a country selecting in an IMF program, $IMF_t = 1$, will be greater conditional on when the wellness of macroeconomics indicates "poor performance" (i.e., $WoM_{t-1} = 0$) compared with "acceptable performance (i.e., $WoM_{t-1} = 1$). To concretize and simplify our thesis, we use a binary-valued indicator of $WoM_{t-1}$; but in reality, it can be multidimensional and continuously valued. We consider any macroeconomics indicator of interest separately and jointly (e.g., economic growth, inflation, balance of payments). Defining the specific indicators and the operationalization of "not acceptable performance" varies depending on the context, which the IMF is well equipped to determine. Following the natural temporality of cause and effect, $WoM_{t-1}$ captures all the relevant macroeconomic parameters before (i.e., $t-1$) countries enter into IMF programs (at time point $t$).

Our sufficientarian-causal framework for selection implies that the propensity to select should be independent of the wellness of individuals. This is indicated in panel *a*, Figure 1, by the absence of a directed arrow from $WoM_{t-1}$ to *IMF*. The probability statement is written as:

$$P(IMF_t = 1|WoM_{t-1} = 0, WoI_{t-1} = 0) = P(IMF_t = 1|WoM_{t-1} = 0)$$

This probability statement implies that the wellness of individuals, $WoI_{t-1} = 0$, does not provide any causal information for self-selecting into programs. In other words, the IMF's decision making on which countries enter into programs should be independent of population health and poverty. However, in many cases, the wellness of individuals (e.g., poor health) causes the wellness of macroeconomics (Banerjee and Duflo 2012; Deaton 2015). This situation is depicted in panel *b* in Figure 1. This is a common situation for many low-income countries, so it is valid to ask whether the IMF should also consider the wellness of individuals (e.g., poverty and health inequality) when determining which countries should enter a program. Based on the spirit of the Articles of Agreement, our suggestion is that the IMF should abstain from these cases; these are situations that fall within the mission of the World Bank and other global organizations.

Once a country has been entered into an IMF program, a sufficientarian approach stipulates that the IMF is expected to design its programs with the causal effects on both the wellness of macroeconomics and individuals taken under consideration. This is depicted in panel *c*, Figure 1, with causal arrows pointing to $WoM_{t+1}$ and $WoI_{t+1}$. For the wellness of macroeconomics, the equation is formulated as:

$$P(WoM_{t+1} = 1 | IMF_t = 1) > P(WoM_{t+1} = 1 | IMF_t = 0)$$

The probability of an acceptable level of wellness of macroeconomics, $WoM_{t+1} = 1$, should be higher after a program (i.e., $IMF_t = 1$) compared with no program (i.e., $IMF_t = 0$). This implies that the IMF is expected to have the intended effect on macroeconomic parameters—the raison d'etre of IMF policies (IMF 2011). As our review reveals, however, what is not adequately calibrated is how these macroeconomic efforts dovetail with the wellness of individuals.

While our causal system in panel *c* highlights the causal effect of IMF on macroeconomics and individuals, panel *d* shows that past wellness levels will, naturally, also have a causal effect on future levels. An empirical analysis determining the extent to which IMF policies are fair would disentangle these causal arrows. We address this issue in the next subsection.

We propose two approaches of formalizing sufficientarian for IMF effects on individuals: lax and stringent. A lax sufficientarian approach would include the IMF's effect on a population as being beneficial or having no effect, on average. This approach implies that although IMF programs are nonadversarial to population health, on average, IMF policies might still emit adverse effects on specific individuals or subgroups. It would not evaluate the potentially considerable variation (e.g., heterogeneity) that IMF policies might produce (Daoud and Johansson 2019). In the lax approach, $WoI_{t+1} = 1$ is an indicator for sufficient population health in the aggregate—for example using the Human Development Index). We define an ethically fair sufficientarian IMF effect as fulfilling the following probability condition:

$$P(WoI_{t+1} = 1|IMF_t = 1) \geq P(WoI_{t+1} = 1|IMF_t = 0)$$

This equation states that populations in IMF programs are either better off after the program or unaffected compared with populations without these programs.

A stringent sufficientarian approach accepts no adverse heterogeneity in an IMF impact: all impact heterogeneity has to be either invariant to the IMF or beneficial for all individuals *i*. This stringent approach follows logically from sufficientarianism as disaggregated. In this case, $WoI_{i,t+1} = 1$ represents individual-level health and poverty outcomes:

$$P(WoI_{i,t+1} = 1|IMF_t = 1) \geq P(WoI_{i,t+1} = 1|IMF_t = 0)$$

This probability implies that each individual is either better off being in an IMF program or unaffected as compared to an individual not in the IMF program.

We refine our definition of $WoI$ based on sufficientarianism—it is any measure of wellness (e.g., malnutrition, education, sanitation) indicating if a person or a group have sufficient resources to satisfy this wellness at a minimum level to avoid deprivation and poverty. For maximizers, more is always better; for sufficientarians, any level beyond this minimum threshold matters little. This distinction is essential. Our stringent sufficientarian framework holds that IMF policies should not be held accountable for lack of improvement at the individual level. The IMF should be held accountable for its policies only when they deteriorate the situation of people, and especially when they push individuals or populations below a minimum threshold. This implies that stringent sufficientarian would claim that IMF policies have unfair effect if, and only if, a program pushes at least one individual, $i$, below the threshold. This logic also implies that if members of a population were above this well-being threshold before an IMF program, but because of the program some or all individuals lost a substantial portion of this well-being but stayed just above the threshold, then stringent sufficientarians would agree that this is not an unfair outcome. It is not unfair because all individuals still have sufficient well-being. However, the level and quality of this "sufficient threshold" varies by context. Following Amartya Sen's agnostic definitions of capabilities (Sen 1992), we emphasize that these thresholds have to be defined for each specific context and done so in public discourse.

Before considering how to empirically measure whether IMF policies fulfill lax or stringent sufficientarianism, we end this section by proposing an amendment to Article V that includes sufficientarian principles. Revising the IMF governing articles in this way would empower the IMF to promote its macroeconomic focus without causing deleterious effects on health and poverty. A reformulation of Article V would reinforce the wording in Article I to "do no harm to national prosperity." A revised Article V, section 3(a) could be:

> The Fund shall adopt policies on the use of its general resources … that will assist members to (i) solve their balance of payments problems, and (ii) ensure that any adverse causes of these policies do not push individuals or populations below the threshold for well-being in a manner consistent with the provisions of this Agreement and that will establish adequate safeguards for the temporary use of the general resources of the Fund.

This reformulation accommodates both stringent and lax sufficientarianism. We remain agnostic to which of these are optimal. At the level of ethical principles, stringent sufficientarianism is more authentic in spirit with an ideal sufficientarianism compared with the lax version. At the empirical level, however, the stringent version requires much more data. In situations with scarce data, it would not be possible to evaluate if IMF policies are fair without strong assumptions about the data-generating process. A stringent approach requires a causal estimate for each individual (e.g., child), while lax sufficientarianism can be routinely captured under the milder assumptions of commonly used causal-inference methods with observational data.

**Empirical identification of lax and stringent sufficientarianism**

To evaluate whether IMF policies comply with sufficientarianism, we have to identify the causal effects of these policies. Because randomized controlled experiments are not possible in researching the IMF, scholars use observational data, but estimating causal effects—empirical identification—from observational data is challenging. The challenges in the literature on IMF policy evaluations pertain to accounting for a country wanting to enter—self-selection—an IMF program for reasons other than wellness of macroeconomics (Vreeland 2007). Empirical identification can be queried by the question: Under what conditions can causal effects of IMF policies on the wellness of individuals and macroeconomics be measured from observational data? There are several methods (i.e., estimators) for doing this (Stubbs et al. 2018), including matching, Heckman selection, instrumental variables (IV), and generalized method of moments (GMM). Each of these are important techniques with different strengths and weaknesses. Our discussion in this section relies on separating what statisticians call *estimands* and *estimators*. Estimands refer to the quantity of interest—in our case, these quantities refer to the causes leading a country to partake in an IMF program and the effects of the policies of those programs on population health. These quantities can be formulated independently of data or models. Conversely, estimators use some data and methods (e.g., IV, matching, GMM) to produce an estimate of this estimand. Our ethical framework pertains first and foremost to sharpening the quantities of interest: the estimand.

Estimating the causal effect relies on blocking or reducing the influence of confounding. Researchers identify vital confounders through causal reasoning that aims to depict the causal mechanisms influencing IMF selection (Imbens and Rubin 2015; Pearl 2009). In this section, we discuss typical causal systems—assumptions encoded in DAGs—affecting the causes and consequences of IMF policies. With the help of these DAGs, scholars can use existing methods (estimators) to estimate the causes and consequences of IMF policies.

The DAGs in Figure 1 facilitate our discussion by assuming that there are no confounders. In these basic causal systems, in which no confounders exist, causal effects can be directly identified from observational data. To evaluate if countries are self-selecting into IMF programs based only on macroeconomic parameters, we estimate the difference as—that is a propensity score model:

$$\delta = P(IMF_t = 1|do(WoM_{t-1} = 1)) - P(IMF_t = 1|do(WoM_{t-1} = 0)).$$

We rely on Pearl's (2009) do-calculus, and use the $do(\cdot)$ operator to signify fixing the causal factor of interest. For example, calculating $do(WoM_{t-1} = 1)$ implies that we intervene on the data by changing all $WoM_{t-1}$ to 1. In general, $do(X = x)$ implies that the causal factor of interest can be fixed and set to $x$. We fix the causal variable of interest to mimic an intervention in the causal system—this fixing is analogous to an intervention in a randomized controlled trial (Pearl 2009). For causal estimation with a binary variable, our estimand is:

$$P(Y|do(X = 1)) - P(Y|do(X = 0))$$

For our IMF selection, if $\delta$ is positive and there are no other causal paths affecting selection (by assumption imprinted in our DAG), then we conclude that IMF selection is fair. To evaluate the impact of IMF policies on the wellness of individuals, we use the equation:

$$\tau = P(WoI_{t+1} = 1|do(IMF_t = 1)) - P(WoI_{t+1} = 1|do(IMF_t = 0))$$

If $\tau$ is positive (i.e., there is a beneficial effect) or zero (i.e., there is no effect), then we conclude that the IMF impact on individuals is fair; otherwise, it is unfair in terms of lax sufficientarianism (we account for the stringent version below). We calculate the effect on the wellness of macroeconomics analogously with the equation:

$$\gamma = P(WoM_{t+1} = 1|do(IMF_t = 1)) - P(WoM_{t+1} = 1|do(IMF_t = 0))$$

and expect a positive $\gamma$. All these situations qualify as an "as-if random" situation.

However, countries self-select into IMF programs for a number of reasons besides macroeconomics, such as political motivation. These other reasons are key confounders and must be accounted for in an empirical identification of causal effects. Figure 2 adds confounding, where the variables $C$ represents one or potentially several confounders. In panel $a$, $C_1$ represents the relevant confounders affecting self-selection. A country's political motivation to implement difficult and unpopular policies is often cited as one of the most important confounders (Åkerström, Daoud, and Johansson 2019; Daoud, Reinsberg, et al. 2019; Daoud and Johansson 2019; Dreher 2009; Stubbs et al. 2018; Vreeland 2007), as are the number of times a country partakes in IMF programs (recidivism). Conditioning or controlling for these measures uses the equation:

$$P(IMF_t|do(WoM_{t-1}), C_1)$$

This enables a calculation of IMF selection fairness, $\delta$. As Figure 2 shows, by conditioning on $C_1$, the influence of this common cause is blocked. We can check if the wellness of individuals is independent of IMF selection conditional on $C_2$ by using the equation

$$P(IMF_t = 1|do(WoM_{t-1}) = 0, WoI_{t-1} = 0, C_2) = P(IMF_t = 1|do(WoM_{t-1} = 0), C_2)$$

Similarly, as seen in panel *b*, Figure 2, we calculate $\gamma$ and $\tau$ by conditioning on $C_3$ and $C_4$, in their respective formulas. Accordingly, identifying a lax sufficientarian approach is empirically measurable when controlling for confounding.

Estimating causal effects by conditioning on confounders depends on available data, and all necessary variables have to be observed. Often, however, important confounders can be unobservable, such as political motivation. In these cases, scholars frequently use an instrumental variable approach. This type of causal situation is depicted in panel *c* in Figure 2, where $Z$ is the instrument. A key assumption of this method is being able to identify an instrument that is uncorrelated with the confounder (i.e., the absence of a causal arrow pointing to the confounder). Countries' "voting patterns in the UN" and "the number of countries already enrolled in IMF programs" have proven to be useful instruments (Lang 2016; Stubbs et al. 2018). Evaluating the degree of sufficientarianism of IMF policies in the presence of unobserved confounding, scholars have to rely on instruments of this sort.

Identifying the causal effect depends on the postulated causal system (DAG). The DAGs in Figures 1 and 2 are options in a number of possible causal systems. As we discussed in the review of empirical studies, the causal pathways connecting IMF to the wellness of individuals (or the wellness of macroeconomics) depends on substantive theory, which vary depending on domain (e.g., economics, sociology, epidemiology). This implies that the probability of wellness outcome, $Y$, conditional on an IMF policy intervention, $do(X = x)$, depends on a specific domain (i.e., causal system) translated into a causa graph, $G_1$. This can be expressed statistically as the probability

$$P_{G_1}(Y|do(X = x), C_1)$$

with subscript $G_1$ to clarify the dependence on the causal system.

Along with the assumptions on the causes and effects of a country self-selecting into an IMF program, a sufficientarian fairness approach is agnostic to the causal system. If there are several competing causal systems, $G_1, G_2, \ldots G_n$, then a reasonable procedure is to sum the estimated effects across the proposed causal systems. This is accomplished via ensembles used in causal approaches in both the computer science (van der Laan, Polley, and Hubbard 2007) and algorithmic fairness literature (Kusner et al. 2017; Loftus et al. 2018).

The estimand of stringent sufficientarianism is defined at the individual level. In do-calculus, the random variable is indexed to denote that each individual is drawn from ones own distribution.

$$P_{G_1}(Y_i|do(X_i = x_i), C_{i1})$$

In the language of potential outcomes, the difference is denoted in treatment effect and indexed for each individual

$$\tau_i = Y_i(1) - Y_i(0)$$

Thus, estimating (that is, identifying) a stringent sufficientarian approach builds on all the same assumptions of a lax sufficientarian approach, but with two additional assumptions. The first is observing each individual $i$ as a specific distribution of outcomes and covariates—$\{Y_i, X_i\}^n$—for all $n$ individuals. The second assumption is that either each individual's *different* causal pathway DAG$_i$ is depicted or that they operate under the same DAG. These two assumptions allow for an estimation of the causal effect for each individual, and an evaluation if stringent sufficientarianism is fulfilled (Kusner et al. 2017; Lillie et al. 2011; Loftus et al. 2018; van and Petersen 2007).

The synthetic control method is one of the few existing estimators that can be used to estimate the estimand $\tau_i$, but this requires high-quality panel data for each individual. Another technique is to impute the Conditional Average Treatment Effect (CATE) using machine-learning techniques (Künzel et al. 2018), as in the equation:

$$\tau(x_i) = E[WoI_{t+1}|do(IMF_t = 1), \boldsymbol{X} = \boldsymbol{x}_i] - E[WoI_{t+1}|do(IMF_t = 0), \boldsymbol{X} = \boldsymbol{x}_i]$$

Here, $\boldsymbol{X}$ is a set of covariates, including confounders. For each individual, this set is drawn from the distribution of the same random variables (i.e., not drawn from different random variables $\boldsymbol{X}_i$ indexed by $i$). This quantity $\tau(x_i)$ allows for an approximation of $\tau_i \approx \tau(x_i)$. This equation implies that each individual's causal effect is estimated on group similarity defined by the covariates $x_i$. Daoud and Johansson (2018) use this method to estimate the impact of IMF programs on child poverty from cross-sectional data, and CATE is the best estimate of individual level treatment effects (Künzel et al. 2018). In reality, this means that the effects on various groups of individuals can be measured, but not for each individual. Even if a stringent sufficientarian approach is measurable only under strong causal assumptions, it remains an important perspective (i.e., estimand) to theorize about the fairness, IMF programs, and individual well-being.

## Conclusions

We have argued that the IMF's Articles of Agreement that govern the organization's mission should be calibrated using the sufficientarian theory. This would allow the IMF to keep its focus on the wellness of macroeconomics but integrate it with the wellness of individuals. The IMF has to account for what causal effects its policies have on populations, both at the aggregate and individual levels. This calibration, however, should not include changes in how the IMF selects countries for its programs, because the wellness of macroeconomics should continue to take precedence.

We based our argument both on distributive ethics and a literature review of the social sciences that empirically evaluates the impact of IMF policies on children's well-being. We used child health and poverty as our case study, and the literature review shows an adverse effect of IMF policies in the majority of these types of cases. The nine studies we reviewed supply evidence that the wellness of macroeconomics and wellness of individuals are, in many cases, at odds. Attempts by the IMF to redesign its policy to not aggravate poverty has proven insufficient. An alternative path, we argued, is to provide an augmented mandate to the IMF that explicitly considers the wellness of individuals. We proposed a revised Article V of the IMF's Articles of Agreement and guiding ethical principles. We showed how, of the four theories discussed above, the IMF's Articles of Agreement could be rearticulated to accommodate a sufficientarian theory, because it is the one most compatible with the IMF mission while simultaneously supporting desired health outcomes. In the context of IMF policies, sufficientarianism holds that a fair allocation of goods allows the IMF and countries in need to formulate economic policies that provide at least the minimum resources required to counter the ill-health and poverty caused specifically by some IMF policies. We showed how the Articles of Agreement could be reformulated qualitatively, and formalized our framework using the language of causal inference and algorithmic fairness.

Our conclusion has several implications. First, to implement our approach, the IMF has to reinforce its staff with experienced social epidemiologists, demographers, and sociologists who can evaluate the causes of population health, social inequality, and individual deprivation beyond income poverty. These experts also possess distinct expertise to evaluate the effects of IMF programs on social, cultural, and individual processes (Woolcock 2009). This new staff would complement IMF economists, who evaluate countries self-selecting into IMF programs and the effects of these programs on the wellness of macroeconomics.

Second, although one practical difficulty in implementing our ethics framework is estimating causal effects from observational data, evaluating lax sufficientarianism is achievable with existing causal-inference methods. It requires estimating the cause (that is, reason) of a specific IMF policy at the population level. While lax sufficientarianism would be easier to implement in practice, stringent sufficientarianism would more aligned to the principles of sufficientarianism. An important connection for stringent sufficientarianism is the recent algorithm for fairness, including in the personalized medicine, literature (Kusner et al. 2017; Lillie et al. 2011). This type of sufficientarianism requires estimating the causal effect of IMF policies at the individual level. Future research can deepen this connection.

Another important area of future research will be developing models that consistently estimate both forward- and backward-looking causality. In policy settings, estimation has to not only work backward in time (ex-post) but also forward in time (ex-ante). The IMF and other similar organizations have to evaluate whether its own policies are fair *before* they implement them. If they are not fair, then they need further calibration before implementation. However, while many traditional causal models can conduct ex-post evaluations, a new method would have to incorporate a portion of causal estimation and causal prediction—beyond pure prediction (Daoud, Kim, and Subramanian 2019). This is an active area of research in machine-learning and causality (Gechter et al. 2019).

Third, our argument supports a substantive shift from economics first to health first (Daniels 1985). It also supports a shift in normative thinking regarding good healthcare policy in favor of putting health (and poverty) first, not justice. Our dual framework keeps the IMF mandate to protect macroeconomic wellness, but not at the cost of the wellness of individuals. When these concerns conflict with one another, tradeoffs have to be empirically grounded to account for the complex interactions between health outcomes and their social determinants (Deaton 2015; Marmot and Wilkinson 2005; McKee et al. 2012).

An objection to our argument is that the IMF's current focus on the wellness of macroeconomics already reflects the importance of how well off individuals are. However, as we argued in Section III, and as many before us have noted, wellness of macroeconomics is not valuable in and of itself (Daoud 2018; Nussbaum 2000; Reddy and Daoud 2020; Sen 1992). Without a functioning economy, population health is expected to take a toll causally, because the economy is only a means to an end.

By contrast, the wellness of individuals is valuable because it directly affects human life; it mitigates significant ill-health and increases health equities among individuals (Krieger 2007). Nevertheless, in addition to mitigating poverty and reducing health inequalities, improving the wellness of individuals has positive effects on other populations in indirect ways. For example, the likelihood that a child has poor health is significantly smaller if the parents are in good health. It is known that individuals in good health contribute more to economic and social development, and more to innovation development that benefits the country.

# References

Adler, Matthew D. 2012. *Well-Being and Fair Distribution: Beyond Cost-Benefit Analysis*. New York: Oxford University Press.

Åkerström, Joakim, Adel Daoud, and Richard Johansson. 2019. "Natural Language Processing in Policy Evaluation: Extracting Policy Conditions from IMF Loan Agreements." *The 22nd Nordic Conference on Computational Linguistics* 5.

Alkire, Sabina. 2015. *Multidimensional Poverty Measurement and Analysis*. First edition. Oxford: Oxford University Press.

Anderson, Elizabeth S. 1999. "What Is the Point of Equality?" *Ethics* 109(2):287–337.

Babb, Sarah. 2005. "The Social Consequences of Structural Adjustment: Recent Evidence and Current Debates." *Annual Review of Sociology* 31:199–222.

Banerjee, Abhijit, and Esther Duflo. 2012. *Poor Economics: A Radical Rethinking of the Way to Fight Global Poverty*. Reprint edition. New York: PublicAffairs.

Barro, Robert J., and Jong-Wha Lee. 2005. "IMF Programs: Who Is Chosen and What Are the Effects?" *Journal of Monetary Economics* 52(7):1245–69.

Beauchamp, Tom L., and James F. Childress. 2012. *Principles of Biomedical Ethics*. 7 edition. New York: Oxford University Press.

Benson, John S. 2001. "The Impact of Privatization on Access in Tanzania." *Social Science & Medicine* 52(12):1903–15.

Bird, Graham, Faryal Qayum, and Dane Rowlands. 2020. "The Effects of IMF Programs on Poverty, Income Inequality and Social Expenditure in Low Income Countries: An Empirical Analysis." *Journal of Economic Policy Reform* 1–19.

Bohoslavsky, Juan Pablo. 2016. "Economic Inequality, Debt Crises and Human Rights." 41:23.

Casal, Paula. 2007. "Why Sufficiency Is Not Enough." *Ethics* 117(2):296–326.

Clegg, Liam. 2014. "Social Spending Targets in IMF Concessional Lending: US Domestic Politics and the Institutional Foundations of Rapid Operational Change." *Review of International Political Economy* 21(3):735–63.

Clements, Benedict, Sanjeev Gupta, and Masahiro Nozaki. 2013. "What Happens to Social Spending in IMF-Supported Programmes?" *Applied Economics* 45(28):4022–33.

Cohen, G. A. 1989. "On the Currency of Egalitarian Justice." *Ethics* 99(4):906–44.

Conklin, Annalijn I., Adel Daoud, Riti Shimkhada, and Ninez A. Ponce. 2018. "The Impact of Rising Food Prices on Obesity in Women: A Longitudinal Analysis of 31 Low-Income and Middle-Income Countries from 2000 to 2014." *International Journal of Obesity* 1.

Copelovitch, Mark S. 2010. *The International Monetary Fund in the Global Economy: Banks, Bonds, and Bailouts*. Cambridge: Cambridge University Press.

Coutts, Adam, Adel Daoud, Ali Fakih, Walid Marrouch, and Bernhard Reinsberg. 2019. "Guns and Butter? Military Expenditure and Health Spending on the Eve of the Arab Spring." *Defence and Peace Economics* 30(2):227–37.

Crisp, Roger. 2003. "Equality, Priority, and Compassion." *Ethics* 113(4):745–63.

Daniels, Norman. 1985. *Just Health Care*. Cambridge Cambridgeshire ; New York: Cambridge University Press.

Daoud, Adel. 2007. "(Quasi)Scarcity and Global Hunger." *Journal of Critical Realism* 6(2):199–225.

Daoud, Adel. 2010. "Robbins and Malthus on Scarcity, Abundance, and Sufficiency." *American Journal of Economics and Sociology* 69(4):1206–29.

Daoud, Adel. 2011. *Scarcity, Abundance, and Sufficiency: Contributions to Social and Econoimc Theory*. Gothenburg: Gothenburg Studies in Sociology, Department of Sociology & Geson Hyltetryck.

Daoud, Adel. 2015. "Quality of Governance, Corruption, and Absolute Child Poverty in India." *Journal of South Asian Development* 10(2):1–20.

Daoud, Adel. 2017. "Synthesizing the Malthusian and Senian Approaches on Scarcity: A Realist Account." *Cambridge Journal of Economics* 42(2):453–76.

Daoud, Adel. 2018. "Unifying Studies of Scarcity, Abundance, and Sufficiency." *Ecological Economics* 147:208–17.

Daoud, Adel, Björn Halleröd, and Debarati Guha-Sapir. 2016. "What Is the Association between Absolute Child Poverty, Poor Governance, and Natural Disasters? A Global Comparison of Some of the Realities of Climate Change." *PLOS ONE* 11(4):e0153296.

Daoud, Adel, and Fredrik Johansson. 2019. "Estimating Treatment Heterogeneity of International Monetary Fund Programs on Child Poverty with Generalized Random Forest." *SocArXiv*.

Daoud, Adel, Rockli Kim, and S. V. Subramanian. 2019. "Predicting Women's Height from Their Socioeconomic Status: A Machine Learning Approach." *Social Science & Medicine* 238:112486.

Daoud, Adel, and Shailen Nandy. 2019. "Implications of the Politics of Caste and Class for Child Poverty in India." *Sociology of Development* 5(4):428–51.

Daoud, Adel, Elias Nosrati, Bernhard Reinsberg, Alexander E. Kentikelenis, Thomas H. Stubbs, and Lawrence P. King. 2017. "Impact of International Monetary Fund Programs on Child Health." *Proceedings of the National Academy of Sciences* 114(25):6492–97.

Daoud, Adel, and Goran Puaca. 2011. "An Organic View of Students' Want Formation: Pragmatic Rationality, Habitus and Reflexivity." *British Journal of Sociology of Education* 32(4):603–22.

Daoud, Adel, and Bernhard Reinsberg. 2018. "Structural Adjustment, State Capacity and Child Health: Evidence from IMF Programmes." *International Journal of Epidemiology* dyy251:1–10.

Daoud, Adel, Bernhard Reinsberg, Alexander E. Kentikelenis, Thomas H. Stubbs, and Lawrence P. King. 2019. "The International Monetary Fund's Interventions in Food and Agriculture: An Analysis of Loans and Conditions." *Food Policy* 83:204–18.

Deaton, Angus. 2015. *The Great Escape: Health, Wealth, and the Origins of Inequality*. Reprint edition. Princeton, NJ: Princeton University Press.

Dorsey, Dale. 2012. "The Basic Minimum: A Welfarist Approach." *Cambridge Core*. Retrieved April 26, 2018 (/core/books/basic-minimum/BEC6C1AEF32D798F5447FCF947D884C5).

Dreher, Axel. 2006. "IMF and Economic Growth: The Effects of Programs, Loans, and Compliance with Conditionality." *World Development* 34(5):769–88.

Dreher, Axel. 2009. "IMF Conditionality: Theory and Evidence." *Public Choice* 141(1–2):233–67.

Dworkin, Ronald. 1981. "What Is Equality? Part 2: Equality of Resources." *Philosophy & Public Affairs* 10(4):283–345.

Feldstein, Martin, and Kathleen Feldstein. 1998. "To IMF: First Do No Harm."

Fleurbaey, Marc. 2015. "Equality versus Priority: How Relevant Is the Distinction?" *Economics & Philosophy* 31(2):203–17.

Gechter, Michael, Cyrus Samii, Rajeev Dehejia, and Cristian Pop-Eleches. 2019. "Evaluating Ex Ante Counterfactual Predictions Using Ex Post Causal Inference." *ArXiv:1806.07016 [Stat]*.

Gordon, D., S. Nandy, C. Pantazis, and S. Pemberton. 2003. *Using Multiple Indicator Cluster Survey (MICS) and Demographic and Health Survey (DHS) Data to Measure Child Poverty*.

Gupta, Sanjeev, Louis Dicks-Mireaux, Ritha Khemani, Calvin McDonald, and Marijn Verhoeven. 2000. "Social Issues in IMF-Supported Programs." *IMF Occasional Papers* 191.

Hajro, Zlata, and Joseph P. Joyce. 2009. "A True Test: Do IMF Programs Hurt the Poor?" *Applied Economics* 41(3):295–306.

Halleröd, Björn, Bo Rothstein, Adel Daoud, and Shailen Nandy. 2013. "Bad Governance and Poor Children: A Comparative Analysis of Government Efficiency and Severe Child Deprivation in 68 Low-and Middle-Income Countries." *World Development* 48:19–31.

Imbens, Guido W., and Donald B. Rubin. 2015. *Causal Inference for Statistics, Social, and Biomedical Sciences: An Introduction*. 1 edition. New York: Cambridge University Press.

IMF. 2011. *Articles of Agreement of the International Monetary Fund*. Washington, DC: International Monetary Fund.

IMF. 2014. "Transcript of the International Monetary and Financial Committee (IMFC) Press Briefing." *International Monetary Fund*. Retrieved (https://www.imf.org/external/np/tr/2014/tr041214b.htm).

IMF. 2016. "Factsheet -- Poverty Reduction Strategy in IMF-Supported Programs." Retrieved August 8, 2016 (www.imf.org).

Inglehart, Ronald, Roberto Foa, Christopher Peterson, and Christian Welzel. 2008. "Development, Freedom, and Rising Happiness: A Global Perspective (1981–2007)." *Perspectives on Psychological Science* 3(4):264–85.

Ismi, Asad. 2004. *Impoverishing a Continent: The World Bank and the IMF in Africa*. Ottawa: Canadian Centre for Policy Alternatives.

Kentikelenis, Alexander, Thomas H. Stubbs, and Lawrence P. King. 2015. "Structural Adjustment and Public Spending on Health: Evidence from IMF Programs in Low-Income Countries." *Social Science & Medicine* 126:169–76.

Kentikelenis, Alexander, Thomas H. Stubbs, and Lawrence P. King. 2016. "IMF Conditionality and Development Policy Space, 1985–2014." *Review of International Political Economy* 23(0):543–82.

Kraamwinkel, Nadine, Hans Ekbrand, Stefania Davia, and Adel Daoud. 2019. "The Influence of Maternal Agency on Severe Child Undernutrition in Conflict-Ridden Nigeria: Modeling Heterogeneous Treatment Effects with Machine Learning." *PLOS ONE* 14(1):e0208937.

Krieger, Nancy. 2007. "Why Epidemiologists Cannot Afford to Ignore Poverty:" *Epidemiology* 18(6):658–63.

Künzel, Sören R., Jasjeet S. Sekhon, Peter J. Bickel, and Bin Yu. 2018. "Meta-Learners for Estimating Heterogeneous Treatment Effects Using Machine Learning." *ArXiv:1706.03461 [Math, Stat]*.

Kusner, Matt J., Joshua Loftus, Chris Russell, and Ricardo Silva. 2017. "Counterfactual Fairness." Pp. 4066–4076 in *Advances in Neural Information Processing Systems 30*, edited by I. Guyon, U. V. Luxburg, S. Bengio, H. Wallach, R. Fergus, S. Vishwanathan, and R. Garnett. Curran Associates, Inc.

van der Laan, Mark J., Eric C. Polley, and Alan E. Hubbard. 2007. "Super Learner." *Statistical Applications in Genetics and Molecular Biology* 6(1).

Lang, Valentin. 2016. *The Economics of the Democratic Deficit: The Effect of IMF Programs on Inequality*. Working paper. 10.11588/heidok.00021875. Heidelberg.

Lillie, Elizabeth O., Bradley Patay, Joel Diamant, Brian Issell, Eric J. Topol, and Nicholas J. Schork. 2011. "The N-of-1 Clinical Trial: The Ultimate Strategy for Individualizing Medicine?" *Personalized Medicine* 8(2):161–73.

Loftus, Joshua R., Chris Russell, Matt J. Kusner, and Ricardo Silva. 2018. "Causal Reasoning for Algorithmic Fairness." *ArXiv:1805.05859 [Cs]*.

Marmot, Michael, and Richard Wilkinson. 2005. *Social Determinants of Health*. OUP Oxford.

McKee, Martin, Marina Karanikolos, P. Belcher, and David Stuckler. 2012. "Austerity: A Failed Experiment on the People of Europe." *Clinical Medicine* 12(4):346–50.

Nandy, Shailen, Adel Daoud, and David Gordon. 2016. "Examining the Changing Profile of Undernutrition in the Context of Food Price Rises and Greater Inequality." *Social Science & Medicine* 149:153–63.

Nussbaum, Martha Craven. 2000. *Women and Human Development: The Capabilities Approach*. Cambridge; New York: Cambridge University Press.

Oliver, Helen C. 2006. "In the Wake of Structural Adjustment Programs: Exploring the Relationship Between Domestic Policies and Health Outcomes in Argentina and Uruguay." *Canadian Journal of Public Health / Revue Canadienne de Sante'e Publique* 97(3):217–21.

Ord, Toby. 2013. "The Moral Imperative toward Cost-Effectiveness in Global Health." 12.

Parfit, Derek. 1997. "Equality and Priority." *Ratio* 10(3):202–221.

Pearl, Judea. 2009. *Causality: Models, Reasoning and Inference*. 2nd edition. Cambridge, U.K. ; New York: Cambridge University Press.

Ponce, Ninez, Riti Shimkhada, Amy Raub, Adel Daoud, Arijit Nandi, Linda Richter, and Jody Heymann. 2017. "The Association of Minimum Wage Change on Child Nutritional Status in LMICs: A Quasi-Experimental Multi-Country Study." *Global Public Health* 13(9):1–15.

Pongou, Roland, Joshua A. Salomon, and Majid Ezzati. 2006. "Health Impacts of Macroeconomic Crises and Policies: Determinants of Variation in Childhood Malnutrition Trends in Cameroon." *International Journal of Epidemiology* 35(3):648–56.

Puaca, Goran, and Adel Daoud. 2011. "Vilja Och Framtid i Frågor Kring Utbildningsval." *Pedagogisk Forskning i Sverige* 16(2):100.

de Rato, Rodrigo. 2006. "Renewing the IMF's Commitment to Low-Income Countries." *IMF*. Retrieved (http://www.imf.org/external/np/speeches/2006/073106.htm).

Rawls, John. 1971. *A Theory of Justice*. Cambridge, Mass.: Belknap Press of Harvard University Press.

Reddy, Sanjay G., and Adel Daoud. 2020. "Entitlements and Capabilities." in *The Cambridge Handbook of the Capability Approach*, edited by E. C. Martinetti, S. Osmani, and M. Qizilbash. Cambridge: Cambridge University Press.

Rickard, Stephanie J., and Teri L. Caraway. 2014. "International Negotiations in the Shadow of National Elections." *International Organization* 68(03):701–20.

Roemer, John E. 1998. *Equality of Opportunity*. Cambridge, Mass.: Harvard University Press.

Sen, Amartya K. 1992. *Inequality Reexamined*. Oxford: Clarendon Press.

Shandra, Carrie L., John M. Shandra, and Bruce London. 2011. "World Bank Structural Adjustment, Water, and Sanitation: A Cross-National Analysis of Child Mortality in Sub-Saharan Africa." *Organization & Environment* 24(2):107–29.

Shandra, Carrie L., John M. Shandra, and Bruce London. 2012. "The International Monetary Fund, Structural Adjustment, and Infant Mortality: A Cross-National Analysis of Sub-Saharan Africa." *Journal of Poverty* 16(2):194–219.

Shandra, John M., Jenna Nobles, Bruce London, and John B. Williamson. 2004. "Dependency, Democracy, and Infant Mortality: A Quantitative, Cross-National Analysis of Less Developed Countries." *Social Science & Medicine* 59(2):321–33.

Shields, Liam. 2012. "The Prospects for Sufficientarianism." *Utilitas* 24(1):101–117.

Stiglitz, Joseph E. 2003. *Globalization and Its Discontents*. 1 edition. W. W. Norton & Company.

Stubbs, Thomas H., Alexander E. Kentikelenis, and Lawrence P. King. 2016. "Catalyzing Aid? The IMF and Donor Behavior in Aid Allocation." *World Development* 78:511–28.

Stubbs, Thomas, Bernhard Reinsberg, Alexander Kentikelenis, and Lawrence King. 2018. "How to Evaluate the Effects of IMF Conditionality." *The Review of International Organizations*.

Stuckler, David, and Sanjay Basu. 2013. *The Body Economic: Why Austerity Kills : Recessions, Budget Battles, and the Politics of Life and Death*.

Summers, Lawrence H., and Lant H. Pritchett. 1993. "The Structural-Adjustment Debate." *The American Economic Review* 83(2):383–89.

Tännsjö, Torbjörn. 1998. *Hedonistic Utilitarianism*. Edinburgh: Edinburgh University Press.

Temkin, Larry S. 1993. *Inequality*. New York: Oxford University Press.

Temkin, Larry S. 2003. "Egalitarianism Defended." *Ethics* 113(4):764–82.

Thomson, Michael, Alexander Kentikelenis, and Thomas Stubbs. 2017. "Structural Adjustment Programmes Adversely Affect Vulnerable Populations: A Systematic-Narrative Review of Their Effect on Child and Maternal Health." *Public Health Reviews* 38:13.

van, der Laan Mark J., and Maya L. Petersen. 2007. "Causal Effect Models for Realistic Individualized Treatment and Intention to Treat Rules." *The International Journal of Biostatistics* 3(1).

Vreeland, James Raymond. 2007. *The International Monetary Fund: Politics of Conditional Lending*. New York, NY: Routledge, Taylor & Francis Group.

Woolcock, Michael. 2009. "Toward a Plurality of Methods in Project Evaluation: A Contextualised Approach to Understanding Impact Trajectories and Efficacy." *Journal of Development Effectiveness* 1(1):1–14.

Ziring, Lawrence, Robert E. Riggs, and Jack C. Plano. 2005. *The United Nations: International Organization and World Politics*. Cengage Learning.

# Figures

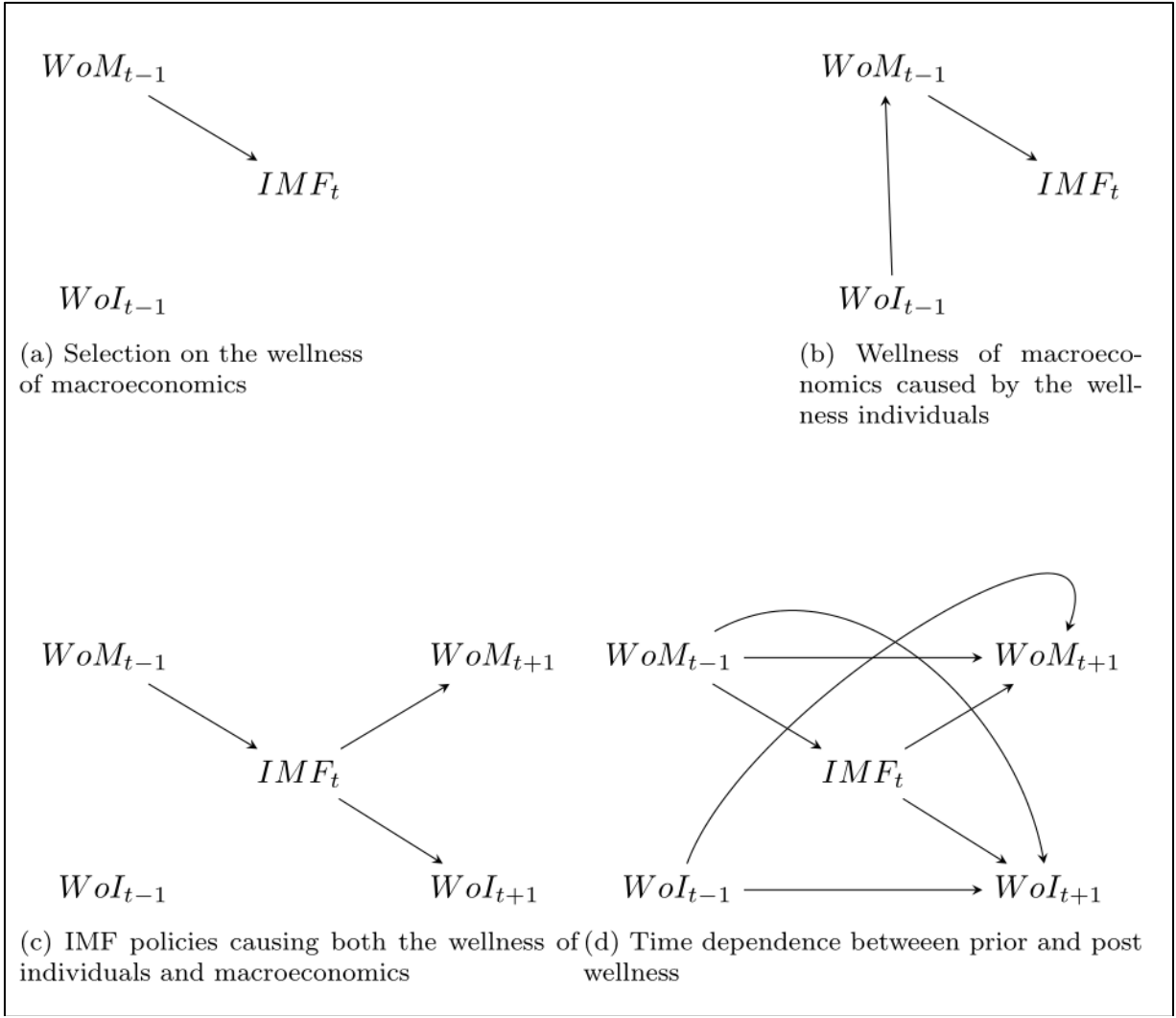

Figure 1: Directed acyclic graphs depicting stylized causal systems of the relationships among IMF policies, the wellness of macroeconomics, and the wellness of individuals.

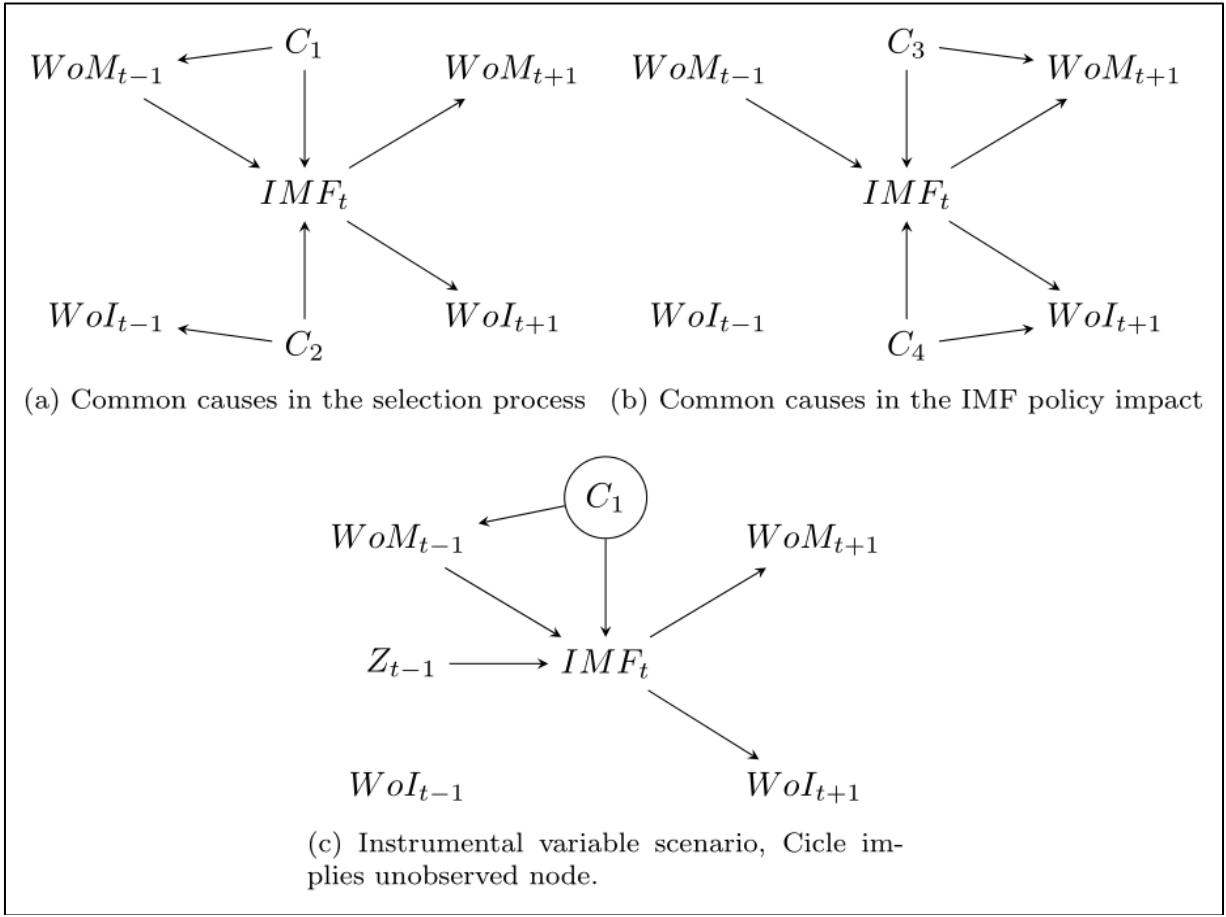

Figure 2: Directed acyclic graphs depicting stylized causal systems with confounding

# Tables

Table 1: the effects of IMF on children's well-being

|  | **Focus** | **Sample** | **Period** | **IMF impact** |
|---|---|---|---|---|
| (Shandra et al. 2012) | Infant mortality | 32 Sub-Saharan countries | 1990-2005 | Adverse |
| (Hajro and Joyce 2009) | Infant mortality | 82 low and middle-income countries | 1985-2000 | Beneficial |
| (Shandra et al. 2004) | Infant mortality | 59 low and middle-income countries | 1980-1997 | Adverse (in interaction with democracy |
| (Oliver 2006) | Infant mortality and child mortality | Argentina & Uruguay | 1980-2000 | Adverse |
| (Pongou et al. 2006) | Child undernutrition | 3510 children in Cameroon | 1991-1998 | Adverse |
| (Daoud et al. 2017) | Child health across five outcomes | 1,941,734 children in 67 low and middle-income countries | 2000 (±5) | Adverse (in moderation with the head of household education) |
| (Daoud and Reinsberg 2018) | Under-five mortality and child vaccination | 128 developing countries | 1980-2014 | Adverse (in IMF public sector policies) |
| (Daoud and Johansson 2019) | Child poverty | 1,941,734 children in 67 low and middle-income countries | 2000 (±5) | Adverse |
| (Bird et al. 2020) | Infant mortality | 48 countries low and middle-income countries | 1990-2015 | Beneficial |

Table 2: Between-population distribution of goods

|  | Outcome 1 | Outcome 2 |
| --- | --- | --- |
| Country A | 100 goods | 75 goods |
| Country B | 50 goods | 70 goods |

Table 3: Within-population distribution of goods

|  | Outcome 1 | Outcome 2 |
| --- | --- | --- |
| Country A | {30, 30, 40}=100 | {10, 10, 55}=75 |
| Country B | {25, 25}=50 | {20, 50}=70 |